\pdfoutput=1

\documentclass[sigconf,natbib=true,anonymous=false]{acmart}

\usepackage[T1]{fontenc}

\usepackage[hang,flushmargin]{footmisc}

\usepackage[T1]{fontenc}    
\usepackage{url}            
\usepackage{booktabs}       
\usepackage{amsfonts}       
\usepackage{nicefrac}       
\usepackage{microtype}      
\usepackage{amsmath}
\usepackage{enumitem}
\usepackage{graphicx}       
\usepackage{caption}
\usepackage{subcaption}
\usepackage{threeparttablex} 

\newcommand{\mytt}[1]{\texttt{\small #1}}

\title{Tevatron: An Efficient and Flexible Toolkit for Dense Retrieval}

\author{Luyu Gao$^1$, Xueguang Ma$^2$, Jimmy Lin$^2$, and Jamie Callan$^1$}

\affiliation{ \vspace{0.1cm}
    $^1$Language Technologies Institute \\
    Carnegie Mellon University \\
  $^2$David R. Cheriton School of Computer Science \\
  University of Waterloo
}

\begin{document}

\begin{abstract}
Recent rapid advancements in deep pre-trained language models and the introductions of large datasets have powered research in embedding-based dense retrieval.
While several good research papers have emerged, many of them come with their own software stacks. 
These stacks are typically optimized for some particular research goals instead of efficiency or code structure.
In this paper, we present Tevatron, a dense retrieval toolkit optimized for efficiency, flexibility, and code simplicity. 
Tevatron provides a standardized pipeline for dense retrieval including text processing, model training, corpus/query encoding, and search. This paper presents an overview of Tevatron and demonstrates its effectiveness and efficiency across several IR and QA data sets.
We also show how Tevatron's flexible design enables easy generalization across datasets, model architectures, and accelerator platforms(GPU/TPU). We believe Tevatron can serve as an effective software foundation for dense retrieval system research including design, modeling, and optimization. 
\end{abstract}
\maketitle

\section{Introduction}
Dense retrieval's popularity in the research community has greatly grown in the past years~\cite{dpr,ance,tct_colbert,rocketqa,condenser}.
By modeling relevance with query-document vector products, dense retrievers can carry out efficient and effective semantic search. 

While the idea of vector-based search is not new, the adoption of deep pre-trained language models as encoder~\cite{bert} has substantially boosted the effectiveness of dense retrieval~\cite{dpr}. 
Meanwhile, similar to other research that relies on deep learning, the success of dense retrieval will not be possible without large data.
Many of recent research works are based on
their own software with specialized support only for specific datasets and models~\cite{dpr, ance}.
We however believe flexible generalization across models and datasets is critical.
Tevatron provides researchers with access to the latest state-of-the-art models and makes it easy for them to start a new research problem on a new dataset.

In our past research on dense retrieval~\cite{condenser, cocondenser, Ma_etal_arXiv2021_DPR}, we have run into several engineering challenges specific to dense systems.
For example, in terms of resources, large corpora and training sets require large CPU memory; accelerator~(GPU/TPU) memory usage also grows with model size. 
While orthogonal to actual research, these engineering problems slow down and constrain researchers, especially those with limited hardware resources.
With Tevatron, we aim at providing a unified solution to common engineering problems.

Tevatron incorporates several popular widely-used 
open-source packages, including datasets~\cite{datasets}, transformers~\cite{transformers} and FAISS~\cite{faiss} respectively as backbone for our data management, neural network modeling and embedding-based retrieval components. 

To accommodate different research needs, we select two deep learning frameworks for Tevatron, Pytorch~\cite{pytorch} and JAX\cite{jax}.
Pytorch's eager execution patterns and intuitive object-oriented design have gained its massive user base in the research community.
On the other hand, JAX, backed by just-in-time~(JIT) XLA compilation, offers smooth transitions across hardware stacks with optimized performance.

The rest of the paper is organized as follows. Section 2 gives an overview of Tevatron. Section 3 demonstrates Tevatron usage and command-line interface. Section 4 shows the experimental results of running Tevatron with various models and datasets.

\section{Toolkit Overview}
Tevatron\footnote{\url{http://tevatron.ai}} is packaged as a Python module available on the Python Package Index.
Tevatron can be installed via \texttt{pip}, as follows:
\smallskip
\begin{quote}
\begin{verbatim}
$ pip install tevatron==0.1.0
\end{verbatim}
\end{quote}

\noindent
In this section, we give an overview of the core components of Tevatron.
We demonstrate how these components respectively support the full pipeline of data preparation, training, encoding, and search.
Code and documentation of Tevatron are available at its website, \texttt{tevatron.ai}.
\subsection{Data Management}
Having data ready to use is a critical preliminary step before training or encoding starts. Data access overhead and constraints could directly affect training/encoding performance. In Tevatron, we adopt the following core design: 1) text data are pre-tokenized before training or encoding happens, 2) keep tokenized data memory-mapped instead of lazy-loaded or in-memory. The former avoids overheads when running sub-word/piece level tokenizers and also reduces data traffic compared to raw text. The latter allows random data access in the training/encoding loop without consuming a large amount of physical memory. 

Tevatron defines two basic raw input format templates for IR and QA context.
As shown in Fig.~\ref{fig_data}, for the IR dataset (e.g. MS MARCO~\cite{msmarco}), we organize a training instance into an anchor query, a list of positive target texts, and a list of negative target texts.
The positive targets are usually human judged and the negative texts are usually non-relevant texts from top results of a baseline retrieval system such as BM25.
\begin{figure}[h]
\begin{small}
\begin{verbatim}
    {
       "query_id": "<query id>",
       "query":    "<query text>",
       "positive_passages": [
         {"docid": "<passage id>",  
          "title": "<passage title>",  
          "text":  "<passage body>"}, ...
       ],
       "negative_passages": [
         {"docid": "<passage id>",
          "title": "<passage title>",
          "text":  "<passage body>"}, ...
       ]
    }
\end{verbatim}
\end{small}
\caption{Tevatron raw data template for IR datasets.}
\label{fig_data}
\end{figure}
The second format (not shown due to space
limits) has an additional \texttt{answers} field for QA tasks (e.g. Natural Question~\cite{natural_questions}), since the positive passages for QA dataset are usually judged by answer exact match~\cite{drqa, dpr}.

Users can pass raw data file pointer and processing specifications to Tevatron's dataset class (\texttt{HFTrainDataset} for training, \texttt{HFQueryDataset} and \texttt{HFCorpusDataset} for encoding) which will perform fast parallel data formatting and tokenization. Processed data is internally represented as a \texttt{datasets.Dataset} object and is stored in Apache Arrow format which can be memory-mapped and randomly accessed by offset.

For researchers who are focusing on building new models, we make a collection of popular open-access datasets self-contained within the Tevatron toolkit. 
For instance, with a single line of command, one can load the training set of MS-MARCO.
Under the hood, Tevatron will first download the raw data set we hosted through Huggingface\footnote{\url{https://huggingface.co/tevatron}}. 
Then it will run the corresponding pre-defined pre-processing script to format and tokenize the downloaded data. 
\subsection{Dense Retrieval Model}

Tevatron's model class \texttt{DenseModel} is a Pytorch \texttt{nn.Module} subclass that defines the deep neural encoder of the dense retriever.
Functionally, it interfaces the underlying Transformer models and provides methods for text encoding and loss computation.
Thanks to duck typing in python, \texttt{DenseModel} class support models in the Huggingface transformer library that return standard base model output. This means new Transformers models can be loaded into Tevatron as soon as they are available in the transformer library.
On the other hand, this helps Tevatron avoid maintaining the transformer codes and reduce code reduplication.
Internally, \texttt{DenseModel} wraps a transformer module and optionally a pooler module which controls how mapping from transformer output tensor to final representations. 
\texttt{DenseModel} class also handles loss computation during training. 
It implements a contrastive loss with in-batch negatives and can perform negative sharing across devices using parallel collective defined in the NCCL library.

\paragraph{JAX Support}
Tevatron has a sub-package \texttt{tevax} that implements core functionality for JAX. 
Following JAX's functional nature~\cite{jax}, \texttt{tevax} is designed with a different philosophy.
We define loss functions that can be \textit{composed} with other JAX transformations.
In practice, they can be combined with Flax models in the transformer library for dense retriever training. 
Two classes \texttt{TiedParams} and \texttt{DualParams} for parameter managing are registered as Pytrees that JAX can differentiate through.
A \texttt{RetrieverTrainState} class manages parameters and model transformations.
With JAX as backend, \texttt{tevax} makes it possible for a single piece of code to run on a single GPU, multiple GPUs, or TPU systems.

\subsection{Trainer}
To complete the dense retriever training setup, we introduce a \texttt{DenseTrainer} which implements miscellaneous training utilities. It controls basic setups such as batch size and the number of training epochs.
When running on multiple GPUs, the trainer will properly set up distributed training and wrap models for gradient reduction.
During training, it will asynchronously load training data to overlap computation and I/O operations. 
At each training step, the trainer turns a batch of loaded data into tensors and passes them to the model.
In this way, the trainer glues the data sets and models together. 

\texttt{DenseTrainer} is a subclass of \texttt{Trainer} in the transformers library.
It inherits a collection of advanced utilities including mixed-precision training and optimizer state sharding.
It is also possible to further subclass \texttt{DenseTrainer} to create unique training behaviors.
Concretely in Tevatron, we implement a subclass \texttt{GCTrainer} which uses gradient caching to support large batch training on memory-limited devices~\cite{gradcache}.

By combining data processor, dense retrieval model, and trainer all together, Tevatron abstracts the training loop of dense retrieval model into the code block shown in Fig.~\ref{fig_train}.
\begin{figure}[t]
\begin{small}
\begin{verbatim}
# initialize model
model = DenseModel.build(model_args)
# initialize dataset
train_dataset = HFTrainDataset(data_args).process()
train_dataset = TrainDataset(data_args, train_dataset)
# initialize trainer
trainer = GCTrainer if training_args.grad_cache else DenseTrainer
trainer = trainer(
    model=model,
    args=training_args,
    train_dataset=train_dataset,
    data_collator=QPCollator(data_args),
)
# start training
trainer.train()
\end{verbatim}
\end{small}
\caption{Training process of Tevatron (Pytorch)}
\label{fig_train}
\vspace{-0.2cm}
\end{figure}

\subsection{Retriever}
The retriever classes in Tevatron build dense retrieval index from text embeddings and execute search over the index.
We use FAISS library~\cite{faiss} as our retriever's backend. 
It implements several efficient indices in C++ and exposes them through Python interfaces.
For users who want the best performance, Tevatron provides a simple class \texttt{BaseFaissIPRetriever} which wraps a flat \texttt{faiss.IndexFlatIP} index for exact search.
Those who want to trade-off between efficiency and effectiveness can use the more powerful \texttt{FaissRetriever} class.
\texttt{FaissRetriever} takes an additional \texttt{index\_spec} string argument in its initialization method and use \texttt{faiss.index\_factory} method to flexibly build the specified index. Users can take advantage of this interface to build approximate search indices like HNSW~\cite{HNSW} or PQ~\cite{PQ}.


\section{Toolkit Usage}
On top of the various core components, Tevatron provides a set of command-line interfaces ~(CLI) to drive the dense retrieval pipeline.
With the flexible design in data and neural model support, one could conduct research of various types without writing code.
In this section, we give an example using Tevatron CLI to run the previously discussed components to learn the model and perform open domain retrieval on Natural Questions~\cite{natural_questions}.
\subsection{Training}

With Tevatron, we are able to replicate the training of DPR model for NQ dataset (see details in section 4.1) by a single command:

\begin{small}
\begin{verbatim}
python -m tevatron.driver.train \
    --do_train \
    --dataset_name Tevatron/wikipedia-nq \
    --model_name_or_path bert-base-uncased \
    --per_device_train_batch_size 128 \
    --train_n_passages 2 \
    --num_train_epochs 40 \
    --learning_rate 1e-5 \
    --fp16 \
    --grad_cache \ 
    --output_dir model_nq
\end{verbatim}
\end{small}

\noindent As introduced in section 2.1, Tevatron will automatically handle the downloading and pre-processing of our self-contained train data \texttt{Tevatron/wikipedia-nq}.
Then the preprocessed dataset and initialized \texttt{DenseModel} will be fed into \texttt{Trainer} class as shown in figure 2.
Since the above command enables the \texttt{grad\_cache} option, it will uses \texttt{GCTrainer} during training.
Here we also enable mix precision training~\cite{fp16} via the \texttt{--fp16} option to improve efficiency.

\subsection{Encoding}
Besides training data, Tevatron also self-contains corresponding corpus data for each dataset.
Again, we simplifies corpus encoding process into a single command:
\begin{small}
\begin{verbatim}
python -m tevatron.driver.encode \
    --output_dir=temp \
    --model_name_or_path model_nq \
    --dataset_name Tevatron/wikipedia-nq-corpus \
    --encoded_save_path corpus_emb_00.pkl \
    --encode_num_shard 20 \
    --encode_shard_index 00 \ 
    --fp16
\end{verbatim}
\end{small}
As encoding the entire corpus within a single process may cost large RAM usage and a long time, Tevatron support encoding the corpus by sharding. 
For example, the above command encodes the first 1/20 split of the entire corpus.
Users can easily run multiple processes for multiple shards in parallel to speed up the encoding process.

\subsection{Retrieval}
By taking query and corpus embeddings, we can run retrieval with following command:
\begin{small}
\begin{verbatim}
python -m tevatron.faiss_retriever \  
    --query_reps query.pkl \  
    --passage_reps corpus_emb_*.pkl \  
    --depth 100 \
    --batch_size -1 \
    --save_text \
    --save_ranking_to result.txt
\end{verbatim}
\end{small}
\noindent where \texttt{--batch\_size} controls the number of queries passed to the FAISS index each search call and -1 will pass all queries in one call. 
Larger batches typically run faster due to better memory access patterns and hardware utilization.
The results will be saved in a text file with each line stores \texttt{query\_id passage\_id score}.

\section{Experiments}
In this section, we demonstrate the system effectiveness and efficiency of Tevatron by running experiments on two common-use collections for QA and IR tasks, Wikipedia and MS MARCO.

\subsection{Comparison with DPR}
The DPR work by Karpukhin et al.~\cite{dpr} is among the first works that show text retrieval using learned dense representations outperforms traditional text retrieval using heuristic sparse representations (e.g. BM25) on open-domain question-answering tasks. 

\begin{table}[h]
\centering\scalebox{0.8}{
\begin{tabular}{llllll}
\hline
         & NQ          & TriviaQA    & SQuAD       & CuratedTrec & WebQuestion \\
\midrule
DPR~\cite{dpr}      & 78.4 / 85.4 & 79.4 / 85.0 & 63.2 / 77.2 & 79.8 / 89.1 & 73.2 / 81.4 \\
Tevatron & 79.8 / 86.9 & 80.2 / 85.5 & 62.3 / 77.0 & 84.0 / 90.7 & 75.4 / 82.9\\
\hline
\end{tabular}}
\vspace{0.2cm}
\caption{Top-20/top-100 retrieval accuracy of DPR model replication on five open-domain QA datasets}
\vspace{-0.6cm}
\label{tab_dpr}
\end{table}

We evaluate the effectiveness of Tevatron by replicating the retrieval results on QA tasks~\cite{natural_questions, triviaqa, squad, curated_trec, web_questions} reported in original DPR work~\cite{dpr}.
We compare the models trained under the "Single" setting defined in the original work where each model is trained by the corresponding individual dataset.
Following the similar hyperparameters setting, we train the models with a learning rate of 1e-5 for 40 epochs with batch size 128.
In Table~\ref{tab_dpr}, except having slightly lower accuracy than the numbers in DPR paper on SQuAD, Tevatron gives even a bit higher top-$k$ accuracy on all other four datasets.
Overall, all top-$k$ accuracy results obtained via the Tevatron pipeline are at the same level of accuracy as original work.
Therefore, we conclude that this is a successful replication, proving that the Tevatron pipeline is effective.

\begin{table}[h]
\centering\scalebox{0.8}{
\begin{tabular}{lccc}
\hline
                   & RAM & GPU memory & Time      \\
\midrule
DPR-repo           & 60G & 20G x 4   & 2.0 hours   \\
Tevatron-default           & 17G & 17G x 4   & 1.5 hours \\
Tevatron-GradCache & 4 G & 15G x 1   & 7.0 hours   \\
Tevatron-TPU       & 10G & --        & 1.0 hours\\
\hline
\end{tabular}}
\vspace{0.2cm}
\caption{Training efficiency comparison between original DPR repo and Tevatron's different setting.}
\vspace{-0.6cm}
\label{tab_efficency}
\end{table}

We demonstrate the efficiency of Tevatron by comparing it with the original DPR repo~\footnote{\url{https://github.com/facebookresearch/DPR}\\ To be clear, the efficiency results are based on the master branch on 2022-02-12} on three dimensions: RAM usage, GPU memory usage, and training time.
The experiments are conducted on a machine with NVIDIA A100 GPUs.
In both DPR-repo and Tevatron-default settings, we train the dense retriever model on 4 GPUs in distributed data-parallel mode of Pytorch.
By comparing the first two rows in Table~\ref{tab_efficency}, we see that Tevatron is more efficient on all three dimensions than the original codebase. Concretely, Tevatron costs 3/4 less RAM, 12G less GPU memory, being 1/4 faster than training using DPR repo.
This means, given the same resources, Tevatron has the potential to support larger training data, larger batch size, and faster training.

The gradient cache feature of Tevatron can further improve the GPU memory efficiency~\cite{gradcache}.
DPR training requires batch size 128 to get the level of retrieval accuracy as reported above.
With the original DPR repo, users cannot train a model with enough batch size if the GPU resource is limited, which will result in a drop in retrieval accuracy. 
Tevatron provides users the option to train dense retrievers using limited GPU resources but keeps the same amount of batch size for each optimization step.
To illustrate this, we conduct experiments with Tevatron-GradCache on a single GPU. 
Tevatron-GradCache trains dense retrievers by splitting the batch of size 128 into sub-batches of size 32. 
Via conducting two round forward steps described in the gradient cache work~\cite{gradcache}, the model update step of Tevatron-GradCache is mathematically equivalent to Tevatron-default.
In the experiment, it costs only 4G RAM and 15G GPU memory, to train a DPR model on NQ dataset with desired batch size.
By reducing the sub-batch size, Tevatron-GradCache can save more GPU memory.

We also evaluated the performance of the training dense retriever model using the Jax backend of Tevatron on a V3-8 TPU VM.
Back in the days when DPR~\cite{dpr} first came out, it cost around a day to train dense retriever on NQ dataset using the initial DPR repo with 8$\times$ Nvidia V100-large GPUs\footnote{This is recorded by DPR authors in the GitHub page.}.
Now it is exciting to see that, such training can be done within one hour with Tevatron.

\subsection{Supervised IR}


To further show the flexibility of the Tevatron toolkit across model architectures and accelerator platforms, we train multiple dense retrieval baselines on MS MARCO passage ranking task with different Transformer backbones from HuggingFace hub\footnote{\url{https://huggingface.co/models}}.
The experiments are conducted on both GPU and TPU platforms.
The models are trained with a learning rate of 5e-6 with batch size 64 for 3 epochs using our self-contained dataset \mytt{Tevatron/msmarco-passage}.

\begin{table}[h]
\centering\scalebox{0.8}{
\begin{tabular}{llrr}
\hline
Model                   & MRR@10 & TPU time & GPU 
time \\
\midrule
1. \mytt{distilbert-base-uncased} & 0.316  &  1.0 hours       &  1.5 hours         \\
2. \mytt{bert-base-uncased}       & 0.322  &  2.0 hours       &  3.0 hours        \\
3. \mytt{co-condenser-marco} & 0.357  &  2.0 hours        &  3.0 hours        \\
4. \mytt{bert-large-uncased}      & 0.327  &  6.0 hours        &   7.5 hours       \\
5. \mytt{roberta-large}          & 0.339  &  6.0 hours       &   7.5 hours    \\
\midrule
6. \mytt{roberta-large} + HN  & 0.361  &  8.0 hours        & 10.0 hours\\
7. \mytt{co-condenser-marco} + HN & 0.382  &  3.0 hours        &  4.0 hours        \\
\bottomrule
\end{tabular}}
\vspace{0.2cm}
\caption{Results of training dense retrievers using Tevatron toolkit with different Transformer model initialization on MS-MSARCO passage ranking task.}
\vspace{-0.8cm}
\label{tab_msmarco}
\end{table}

In Table~\ref{tab_msmarco}, we show MRR@10 for each model and training time on 4$\times$ A100 GPU and V3-8 TPU.
The model backbones we choose varying across:
\begin{itemize}
    \item model size: row(1), row(2) and row(4) are models in BERT family with different size \{distil, base, large\}.
    \item model type: row(4), row(5) are models in same level of size with different backbone structure \{bert, roberta\}.
    \item model parameters: row(2), row(3) are same model backbone but the later one is further pre-trained from row(2), i.e. \{original, fine-tuned\}.
\end{itemize}

\noindent For all the variants of model initialization, Tevatron can train dense retrievers effectively and efficiently on different platforms with the Tevatron CLI commands.

Finally, we evaluated two models trained with hard negative mining\cite{cocondenser} mined with Tevatron retriever.
We craft the augmented training data by combining the hard negative passages mined using the first round dense retriever model with the original training dataset.
Then we retrain the models using the hard negative augmented data.
The last two rows in Table~\ref{tab_msmarco} show that by augmenting training data with hard negative, we can further improve the effectiveness of the dense retriever model.
We also demonstrate Tevatron can replicate the state-of-the-art co-Condenser retriever\cite{cocondenser} on MS MARCO passage ranking.

\subsection{Cross-lingual Retrieval}
The success of dense retrieval also drives research in multilingual retrieval~\cite{xorqa, mrtydi, tydiqa}.
We additionally show our Tevatron toolkit can generalize to multilingual retrieval tasks by replicating the dense retrieval baseline reported in the XOR-Retrieve task~\cite{xorqa}.

\begin{table}[h]
\centering\scalebox{0.8}{
\begin{tabular}{lllllllll}
\hline
         & Ar   & Bn   & Fi   & Ja   & Ko   & Ru   & Te   & avg  \\
\midrule
mDPR     & 50.4 & 57.7 & 58.9 & 37.3 & 42.8 & 44.0 & 44.9 & 48.0 \\
Tevatron & 50.5 & 64.1 & 57.3 & 41.9 & 60.4 & 48.5 & 58.4 & 54.4 \\
\bottomrule
\end{tabular}}
\vspace{0.2cm}
\caption{Recall@5kt of mDPR replication on dev set of \texttt{XOR-RETRIEVE} task with Tevatron}
\vspace{-0.6cm}
\label{tab_mdpr}
\end{table}

\noindent We train dense retriever with Tevatron that encode seven languages queries and English corpus into the same embedding space.
Such a method can conduct retrieval in a single stage without addition need for translation.
Results in Table~\ref{tab_mdpr} show that the baseline model replicated with Tevatron gaining on average 6 points over original baseline results on seven languages.  

\section{Conclusion}
This paper introduces Tevatron, an efficient and flexible toolkit for training and running dense retrievers with Transformers.
The toolkit has a modularized design for easy research exploration and a set of command-line interfaces for fast development and evaluation.
Our experiments show that Tevatron can be used to train dense retrieval models effectively and efficiently.
The flexible and generalizable functionalities provide IR community convenience in future dense retrieval research.

\section*{Acknowledge}
We would like to thank Google’s TPU Research Cloud (TRC) for access to Cloud TPUs and Compute Canada for access to GPU clusters.



\bibliographystyle{ACM-Reference-Format}
\bibliography{ref}

\end{document}